\documentclass[conference]{IEEEtran}
\IEEEoverridecommandlockouts
\usepackage{cite}
\usepackage{amsmath,amssymb,amsfonts}
\usepackage{amsthm}
\usepackage{algorithmic}
\usepackage{algorithm}
\usepackage{graphicx}
\usepackage{textcomp}
\usepackage{xcolor}
\usepackage{todonotes}
\usepackage{url}
\usepackage{cleveref}
\usepackage{bm}
\usepackage[braket, qm]{qcircuit}
\usepackage{mathtools}
\usepackage{subcaption}
\usepackage{pdflscape}

\DeclareMathOperator*{\argmin}{argmin}
\newcommand{\abs}[1]{\left\lvert#1\right\rvert}

\crefformat{equation}{(#2#1#3)}
\Crefname{figure}{Fig.}{Fig.}

\newtheorem{theorem}{Theorem}

\newtheorem{proposition}[theorem]{Proposition}

\theoremstyle{definition}
\newtheorem{definition}[theorem]{Definition}
\newtheorem{example}[theorem]{Example}
\newtheorem{remark}[theorem]{Remark}
\newtheorem*{proof_prop}{Proof of Proposition \ref{prop}}

\DeclareMathOperator*{\Expect}{\mathbb{E}}

\def\BibTeX{{\rm B\kern-.05em{\sc i\kern-.025em b}\kern-.08em
    T\kern-.1667em\lower.7ex\hbox{E}\kern-.125emX}}
\begin{document}

\title{Quantum Architecture Search for\\
Quantum Monte Carlo Integration via\\
Conditional Parameterized Circuits\\
with Application to Finance
\thanks{This work was supported by the project AnQuC-3 of the Competence Center Quantum Computing Rhineland-Palatinate (Germany).}
}

\author{\IEEEauthorblockN{Mark-Oliver Wolf}
\IEEEauthorblockA{\textit{Department of Financial Mathematics} \\
\textit{Fraunhofer ITWM}\\
Kaiserslautern, Germany \\
mark-oliver.wolf@itwm.fraunhofer.de\\
0000-0002-3698-9266}
\and
\IEEEauthorblockN{Tom Ewen}
\IEEEauthorblockA{\textit{Department of Financial Mathematics} \\
\textit{Fraunhofer ITWM}\\
Kaiserslautern, Germany \\
tom.ewen@itwm.fraunhofer.de}
\and
\IEEEauthorblockN{Ivica Turkalj}
\IEEEauthorblockA{\textit{Department of Financial Mathematics} \\
\textit{Fraunhofer ITWM}\\
Kaiserslautern, Germany  \\
ivica.turkalj@itwm.fraunhofer.de}
}


\maketitle

\begin{abstract}
    Classical Monte Carlo algorithms can theoretically be sped up on a quantum computer by employing amplitude estimation (AE). To realize this, an efficient implementation of state-dependent functions is crucial. We develop a straightforward approach based on pretraining parameterized quantum circuits, and show how they can be transformed into their conditional variant, making them usable as a subroutine in an AE algorithm. To identify a suitable circuit, we propose a genetic optimization approach that combines variable ansatzes and data encoding. We apply our algorithm to the problem of pricing financial derivatives. At the expense of a costly pretraining process, this results in a quantum circuit implementing the derivatives' payoff function more efficiently than previously existing quantum algorithms. In particular, we compare the performance for European vanilla and basket options.
\end{abstract}


\section{Introduction}

The potential for revenue and market disruption of quantum computing is enormous for the financial sector, with estimates going as high as \$5 billion \cite{IBMexpertInsights2019}.
Showing a practical quantum advantage is crucial to attain these estimates. 
In gate-based quantum computing, we are currently in the era of Noisy Intermediate Scale-Quantum (NISQ) devices \cite{Preskill2018quantumcomputingin}.
These are quantum devices with high error rates, therefore requiring potential quantum circuits to be extremely efficient in their resources such that the needed error mitigation schemes may be applied \cite{bauckhage_quantum_2022}.

In finance, the pricing and risk management of assets, such as stocks, bonds, commodities, currencies and their derivatives is used for a variety of applications and regulations. Depending on the studied instrument, computing the necessary values on classical computers is complex and time consuming.
There have been a variety of publications studying how the application of quantum computers may speed this up tremendously \cite{orus_quantum_2019, herman_survey_2022}.
A promising approach is the substitution of integration using classical Monte Carlo algorithms with its quantum counterpart, generally referred to as quantum Monte Carlo integration (QMCI) \cite{intallura_survey_2023}.
A variety of QMCI methods use a quantum algorithm called Amplitude Estimation (AE) \cite{brassard_quantum_2000}, leading to a quadratic speed-up of the rate of convergence in the number of simulations \cite{rebentrost_quantum_2018,Stamatopoulos2020optionpricingusing}.
One crucial aspect of these simulations is the implementation of the respective functional dependencies on the quantum device \cite{vazquez_efficient_2020}. Similar to classical simulations, these functions have to be applied multiple times in the estimation algorithm.

In this work, we present a novel method to substantially reduce the required quantum resources for this function implementation, 
at the cost of a computationally expensive pre-training on a classical device.
Employing methods from the thriving area of quantum machine learning (QML) \cite{Biamonte_2017}, 
we automatically generate the structure for a parameterized quantum circuit to implement payoffs of a financial derivative.
Further, we introduce a technique based on substituting encoding circuits by controlled unitaries. 
We use the trained circuits with a varying number of qubits, which for example can be used for increasingly fine discretizations of the payoff function.
This has two main advantages. First, one does not have to design a suitable circuit by hand, which, depending on the payoff, may not be trivial.
Second, the resulting overhead for reusing a parameterized circuit scales only linearly with
the number of qubits $n$ used for the discretization,
while implementing a function that can map $2^n$ different states to their respective values.
Hence, easing the cost of simulations on a quantum computer.
Interestingly, this approach evades the typical problem of data loading efficiency that is present in various QML applications \cite{schuld2021machine}. 
Instead, we encode specific values as efficiently controlled operators.
The result is a pre-trained circuit that can be thought of as a conditional variant of the initial parameterized quantum circuit.

The article is structured as follows. The theoretical foundation of our algorithm is presented in \Cref{sec:cpqc}.
We prove that the desired equality of expectations directly follows from an equality of individual encodings.
This enables us to find an efficient conditional variant for circuits that use a specific rotational encoding.
In \Cref{sec:implementation_of_genetic_pqcs}, we focus on finding a suitable, unconditonal circuit which we may later use to construct its conditional variant. 
Thereby, we present a genetic approach to automated circuit design, using prior results from Variable Ansatzes (VAns) \cite{bilkisSemiagnosticAnsatzVariable2023} present in Variational Quantum Eigensolvers (VQE). 
This genetic approach is tailored for current NISQ devices and can be set to specifically use hardware-native quantum gates.
In \Cref{sec:application_to_options}, we apply and compare the conditional parameterized quantum circuit to current state-of-the-art quantum algorithms for the pricing of financial derivatives, namely European vanilla options, as well as basket options with a fixed and variable weight.
Surprisingly, the variable weight basket option only encodes the weight a single time at the start of the circuit.
We conclude with a short summary and prospective future research in \Cref{sec:conclusion}.

\section{Conditional Parameterized Quantum Circuits}\label{sec:cpqc}

We give a short overview why AE has the potential for a speed up compared to classical Monte-Carlo methods.
Generally, they are used to calculate a variety of statistics for a given random variable $X$, respectively functionals $f(X)$.
In quantitative finance, we are often interested in the expectation of $f(X)$.
Unfortunately, knowing the distribution of $X$ and the function $f$ does not lead to us necessarily knowing anything about the distribution of $f(X)$.
Especially, \(\Expect \left[ f(X) \right]\) is in general not the same as \(f\left(\Expect \left[X\right]\right)\).
For many applications, there is no closed formula known for this expectation, so we have to estimate it by
\begin{align}\label{eq:mean_approx}
    \Expect \left[ f(X) \right]
    \approx
    \frac{1}{M} \sum_{i=1}^M f(X_i),
\end{align}
where $M$ is the number of realizations $X_i$ of $X$. This sample mean estimator on the right-hand side converges with a rate of $O(1/\sqrt{N})$ by the Central Limit Theorem.
For AE, we have to find a unitary operator $\mathcal{A}$ that constructs a quantum state with
\begin{align*}
    \mathcal{A} \ket{0}_{c} \ket{0}
    =
    a \ket{\varphi_0}_c \ket{0} + b \ket{\varphi_1}_c \ket{1},
\end{align*}
for which we can now estimate the needed probability $|b|^2$ with a convergence of $O(1/N)$ w.r.t.\ the applications of $\mathcal{A}$. In our notation, the index $c$ indicates the corresponding qubit register. 
By discretizing the range of $X$, we can find an operator \cite{Stamatopoulos2020optionpricingusing, Woerner_2019} such that
\begin{align}\label{eq:AE_for_MC}
    |b|^2
    \approx
    \sum_{\omega \in \Omega} p_\omega f(X_\omega)
    =
    \Expect \left[ f(X) \right],
\end{align}
where $\Omega$ is a discretized state space of $X$ with probabilities $p_\omega$ and $X_\omega=X(\omega)$. This results in a theoretical speed up compared to classical algorithms.

Equation $\eqref{eq:AE_for_MC}$ can be realized by loading a distribution, implemented by the unitary $P$, on a control (or state) register of $n$ qubits $\ket{0}_c$ and applying a function $f$, implemented by the unitary $F$, using an additional target qubit $\ket{0}_t$ depending on this control register:
\begin{align}\label{eq:FP_application_state}
    F \left( P \otimes I_{1} \right) \ket{0}_c \ket{0}_t
    =
    \ket{\varphi_0}_c \ket{0}_t
    +
    \sum_{k=0}^{2^n-1} \sqrt{p_k} \ket{k}_c \sqrt{f(k)} \ket{1}_t.
\end{align}
In our notation, $\ket{k}$ for $k\in \mathbb{N}$ always refers to the computational basis state of the respective binary representation, and $\ket{\varphi}$ without additional comments are arbitrary quantum states.
It is easier to find a quantum circuit for $P$ for sufficiently smooth distributions \cite{grover2002creating}. But in general, the problem is nontrivial and referred to as Quantum State Preparation \cite{Araujo_2021}.

However, implementing the unitary $F$ depends on the task at hand, and might introduce significant circuit depth, even if the underlying distribution is simple enough. 
In the following, we present a novel approach to accomplish this with smaller circuits.

\subsection{Parameterized Quantum Circuits}

One way to grasp hard-to-implement functions is to use parameterized quantum circuits (PQC)
as function approximators. 
It has been shown that a sufficiently regular function can be approximated 
with arbitrary accuracy by a certain type of PQC.
These results are based on the representation of a PQC and its 
expected measurement result by truncated Fourier series 
\cite{schuld2021effect, yu2022power}.
To achieve a good approximation quality, PQCs with high depth are needed,
making it problematic to run them on NISQ devices.
Therefore, we also consider genetic algorithms, which can be designed to result in shallower PQCs.

A widely used form of PQCs, which we also consider in this paper, is given by
\begin{align}\label{eq:pqc}
    U(\bm{x},\bm{\theta}) := 
    \prod_{l=1}^{L} S_l(\bm{x}) W_l(\bm{\theta}^l),
\end{align}
where $\bm{x} = (x_1,\ldots, x_\mathcal{K}) \in \mathbb{R}^\mathcal{K}$ is a feature vector
and for each $l = 1, \ldots, L$ we have that 
$\bm{\theta}^l = (\theta_1, \ldots, \theta_{Q_l}) \in \mathbb{R}^{Q_l}$
is a vector of parameters and
\begin{align} \label{eq:blocks}
    S_l(x) := \prod_{k=1}^{\mathcal{K}} T_{l,k} \cdot e^{-ix_k H_{l,k}},
    \quad
    W_l(\bm{\theta^l}) := \prod_{q=1}^{Q_l} V_{l,q} \cdot e^{-i\theta_q G_{l,q}}
\end{align}
are quantum circuits that we denote as encoding and parameterized blocks, respectively.
Here, $T_{l,k}, V_{l,q}$ are unitary and
$H_{l,k}, G_{l,q}$ are hermitian matrices.
We use $\bm{\theta} := \bm{\theta}^1 \cdots
\bm{\theta}^L \in \mathbb{R}^Q$ 
to abbreviate the concatenation of all parameters, where 
$Q :=Q_1+ \ldots + Q_L$.

We define the quantum model for $U(\bm{x},\bm{\theta})$ to be the function
\begin{align*}
    f_{U}: \mathbb{R}^\mathcal{K} \times \mathbb{R}^Q & \longrightarrow \mathbb{R}, \\
    (\bm{x}, \bm{\theta}) & \longmapsto 
    \bra{0}U(\bm{x},\bm{\theta})^{\dagger} \mathcal{M}
    U(\bm{x},\bm{\theta})\ket{0},
\end{align*}
where $\mathcal{M}$ is some observable. In QML, a typical choice is the $\sigma_Z$ observable on one or more qubits, which we use as well if nothing else is specified.

For a given set of feature vectors 
$\mathcal{X} = \{ \bm{x}^1, \ldots, \bm{x}^J \} \subset \mathbb{R}^\mathcal{K}$
and corresponding labels $\mathcal{Y} = \{y_1, \ldots, y_J\} \subset \mathbb{R}$,
the quantum model $f_{U}$ is trained by minimizing over $\bm{\theta}$ while
averaging the distance between the quantum model of the data point and the corresponding labels over $\mathcal{X}$,
\begin{align*}
    \argmin_{\bm{\theta} \in \mathbb{R}^Q}
    \abs{f_U(\bm{x}^j,\bm{\theta}) - y_j}^2.
\end{align*}

We refer to the whole process of constructing $U(\bm{x},\bm{\theta})$
and finding a minimizing set of parameters as a Variational Quantum Algorithm (VQA)
\cite{Cerezo2020VariationalQA}, \cite{peruzzo2014}.

\subsection{Conditional Variant to Implement State-Dependency}

We are now introducing the main theoretical concept of this article, the conditional parameterized quantum circuit (CPQC).
As mentioned in the introduction, one benefit of the CPQC approach is that one 
can calculate the weighted average of a function (approximated by a PQC) by evaluating a 
circuit on a suitable superposition of computational basis states.
Here, we describe the idea in a more general context, where different distributions
are loaded into different, not necessarily same sized, registers simultaneously.

Let $\mathcal{X} = \{ \bm{x}_1, \ldots, \bm{x}_{2^n} \} \subset \mathbb{R}^\mathcal{K}$
be a data set. 
For each dimension $k$ of the feature vector, we assume a quantum register with $n_k$ qubits. It should be emphasized again that we do not need these registers to be the same size.
This means that if they are used for implementing discretizations of random variables, the chosen resolutions are allowed to differ.
These registers are indexed by $r_1, \ldots, r_\mathcal{K}$ and called control registers.
We denote with $n = \sum_{k=1}^{\mathcal{K}} n_k$ the total number of qubits used for the (entire) control register.
Associated with each control register is a distribution $\mathcal{P}_k$
with $|p_{k_1}|^2, \ldots, |p_{k_{2^{n_k}}}|^2, p_{k} \in \mathbb{C}$ denoting the probabilities of the corresponding basis states.
Further, we assume there to be an additional target register consisting of $m$ qubits, indexed by $t$.

Let $U$ be a unitary acting on the target register and let 
$c = \{i_1,\ldots,i_k\} \subseteq \{1,\ldots,n\}$ be a subset indicating a choice
of qubits.
As usual \cite{nielsen_chuang_2010}, 
the controlled operation $\Lambda(c,U)$ is defined 
by its action on the computational basis,
\begin{align*}
    \Lambda(c,U) \ket{b_1 \ldots b_n}_c \ket{j}_t
    :=
    \ket{b_1 \ldots b_n}_c U^{b_{i_1}\cdots b_{i_k}}\ket{j}_t,
\end{align*}
where $b_1, \ldots, b_n \in \{0,1\}$ and $j=0,\ldots,2^{m-1}$.
In other words, the unitary $U$ is applied exactly if all $b_{i} = 1$ for qubits $i \in c$, otherwise we apply the identity.
We use the notation $\Lambda^{(r)}(c,U)$ for the control operation acting on $n+m$ qubits, to indicate that the control qubits $c$ are taken from the register $r$ and that $\Lambda^{(r)}(c,U)$ acts trivially on all remaining registers.
For some index set $D$, we call a product of control operations,
\begin{align*}
    \Lambda := \prod_{k=1}^{\mathcal{K}}\prod_{d \in D}\Lambda^{(r_k)}(c_d,U_d)
\end{align*}
a control circuit. 

We now define the main concept of this paper.

\begin{definition}(Conditional PQC) \label{def:cpqc}
    \newline
    Let $U:=U(x, \bm{\theta}) = 
    \prod_{l=1}^{L} S_l(\bm{x}) W_l(\bm{\theta}^l)$
    be a PQC. A conditional parameterized quantum circuit (CPQC)
    associated to $U$ is a unitary of the form
    \begin{align*}
        C_n(U) := \prod_{l=1}^{L} \Lambda_l \cdot (I_{n} \otimes W_l(\bm{\theta}^l)),
    \end{align*}
    where $\Lambda_l$ is a control circuit for each $l=1, \ldots, L$.
\end{definition}

The idea is that every entry of a feature vector $\bm{x}$ enters the CPQC through 
its designated control 
register via basis encoding. 
To make this idea more precise, we presume a mapping 
$b: \mathbb{R} \longrightarrow \{0,1\}^n$ that specifies a 
translation of entries of a feature vector into basis states.

For each $k=1,\ldots,\mathcal{K}$, the distribution $\mathcal{P}_k$ is loaded into the
control register $r_k$ via a unitary $P_k$ defined through the relation 
\begin{align*}
    P_k\ket{0}_{r_k} = \sum_{i=0}^{2^{n_k}-1} \sqrt{p_{k_i}}\ket{b(\bm{x}_i^k)}_{r_k}.
\end{align*}
The unitary $P:=P_1 \otimes \cdots \otimes P_\mathcal{K}$ can then be interpreted as
loading the joint distribution of independent random variables into the
whole $n$-qubit control register.

Similar to PQCs, we define a quantum model for a CPQC to be the function
$f_{C_n(U),P}$ defined by
\begin{align*}
    \bm{\theta} \longmapsto 
    \bra{0}(P \otimes I_m)^{\dagger}C_n(U)^{\dagger} \mathcal{M} C_n(U)(P \otimes I_m)\ket{0}.
\end{align*}
The observable $\mathcal{M}$ is always chosen as $I_n \otimes M$, if $M$ 
is the observable of the associated PQC.

\begin{remark}(Advantages of CPQC)
    \newline
    The ensuing question is how to construct $C_n(U)$ such that
    \begin{align}\label{eq:cpqc_goal}
        \sum_{i=0}^{2^n-1} p_i f_U(\bm{x}^i,\bm{\theta}) = f_{C_n(U),P}(\bm{\theta}),
    \end{align}
    where the coefficients $p_i$ are the values of the probability mass function of 
    the joint distribution. Equation \cref{eq:cpqc_goal} states that,
    using $C_n(U)$, the classically expensive approximation in \cref{eq:mean_approx} can be 
    performed on a quantum computer by applying the same PQC $U$, without changing the ansatz when the amount of 
    data points increases.
\end{remark}

The following proposition gives a sufficient condition for the satisfiability
of the above equality.

\begin{proposition}(Equality of encodings)\label{prop}
    \newline
    Let $U$ be a PQC and $C_n(U)$ a corresponding CPQC. Then
    \Cref{eq:cpqc_goal} is satisfied,
    if
    \begin{align*}
        (I_{n} \otimes S_l(\bm{x}))\bigotimes_{k}\ket{b(\bm{x}_k)}_{r_k}\ket{\varphi}_t
        =
        \Lambda_l \bigotimes_{k} \ket{b(\bm{x}_k)}_{r_k}\ket{\varphi}_t
    \end{align*}
    for all $\bm{x} \in \mathcal{X}$ and all $\ket{\varphi}_t$ in the space of the 
    target register.
\end{proposition}

Observe that the circuit on the left-hand side depends on $\bm{x}$, whereas
the circuit on the right-hand side does not.
The condition in the proposition means that one replaces the encoding circuits
$S_l(\bm{x})$ by control circuits $\Lambda_l$ that receive their input via the control register.
Both encoding procedures are chosen to produce the same quantum state for the same computational basis states.
\Cref{prop} ensures that this equality of basis states translates to the case of the control register being in a superposition, without possible phase kickbacks leading to additional error terms due to the repeated controlling procedure.
The proof is based on the linearity of the expected value and can be carried out
using standard rules for the kronecker product.
It is given in the appendix.

In the next section, we illustrate the above proposition by concrete examples.

\subsection{Efficient CPQCs}

For brevity, we only consider examples of one-dimensional features in this section, i.e., $\mathcal{K}=1$.

\begin{example}
    \ \newline
    The following example shows how it is always possible to construct a CPQC in a trivial way that reproduces the result of a given PQC.

    Let $\mathcal{X} = \{ x_1, \ldots, x_{2^n} \}$ be our data. 
    For each $x \in \mathcal{X}$, let $c(x) \subseteq \{1,\ldots,n\}$ be 
    the set of indices representing the positions of $1$ in $b(x)$.
    Let $U(\bm{x},\bm{\theta}) = 
    \prod_{l=1}^{L} S_l(\bm{x}) W_l(\bm{\theta}^l)$
    be a PQC.
    For each $l=1,\ldots,L$, we define a control circuit as follows:
    \begin{align*}
        \Lambda_l(n) := \prod_{x \in \mathcal{X}} \Lambda(c(x),S_l(x)).
    \end{align*}
    This circuit fulfills the requirement of \Cref{prop}.
    Note however, that for arbitrary data $\mathcal{X}$ and encoding blocks $S_l(x)$ constructing $\Lambda_l(n)$ needs $\Omega(2^{n+1})$ elementary (CNOT and one-qubit one-parameter rotational) gates \cite{mottonen2004transformation}.
\end{example}

We see that a concept of efficiency is needed to find a distinction of \textit{bad} and \textit{good} control circuits $\Lambda$. Hence, we introduce the following definition.

\begin{definition}(Efficiency)\label{def:effciency}
    \newline
    Let $U$ be a PQC. Assume we have a procedure that generates a CPQC,
    $C_n(U)$, for every $n \in \mathbb{N}$.
    We call this construction efficient, if the minimum number of elementary gates needed to construct $C_n(U)$ scales linearly with $n$, i.e., is in $O(n)$.
\end{definition}

We proceed with a small example showcasing an efficient CPQC construction, and afterwards generalize the concept for arbitrarily sized control registers and PQCs that have direct encodings via rotational gates.

\begin{example}\label{ex:3qubits}
    \ \newline
    We choose $n=3$ control qubits and $m=1$ target qubit.
    Let $\mathcal{X}$ and $b: \mathcal{X} \longrightarrow \{0,1\}^3$ be given by the 
    following table:
    \begin{center}
        \scalebox{0.9}{
        \begin{tabular}{|c|c|c|c|c|c|c|c|c|}
            Index $i$ & 0 & 1 & 2 & 3 & 4 & 5 & 6 & 7 \\
            \hline
            $x_i \in \mathcal{X}$ & 0 & 0.1 & 0.2 & 0.3 & 0.4 & 0.5 & 0.6 & 0.7 \\
            \hline
            $b(x_i)$ & 000 & 001 & 010 & 011 & 100 & 101 & 110 & 111 
        \end{tabular}
        }
    \end{center}

   The PQC consists of one layer only and the encoding circuit used is 
   $S_1(x) = R_{\mathrm{Y}}(x)$. The parameterized circuits $W(\bm{\theta})$ 
   and the control circuits $\Lambda_l$ used to construct the CPQC are shown
   in \Cref{fig:example_CVQC}. 
   It is evident that, with the data and encoding chosen as above, the requirement
   of \Cref{prop} is satisfied.

   \begin{figure}
    \centering
    \subcaptionbox{PQC}{
        \scalebox{1}{
        \Qcircuit @C=1em @R=.7em {
            & \gate{W_2(\bm{\theta}^2)} \ar@{--}[]+<2.4em,1.3em>;[]+<2.4em,-1.3em> 
            & \gate{R_{\mathrm{Y}}(x)} \ar@{--}[]+<2.2em,1.3em>;[]+<2.2em,-1.3em> 
            & \gate{W_1(\bm{\theta}^1)}
            & \qw
        }}
    }
    \par\bigskip
    \subcaptionbox{CPQC}{
        \scalebox{0.9}{
        \Qcircuit @C=1em @R=.7em {
            & \qw \ar@{--}[]+<2.4em,1em>;[ddd]+<2.4em,-1em>
            & \ctrl{3} & \qw 
            & \qw \ar@{--}[]+<2.5em,1em>;[ddd]+<2.5em,-1em> 
            & \qw & \qw 
            \\
            & \qw & \qw & \ctrl{2} & \qw & \qw & \qw
            \\
            & \qw & \qw & \qw & \ctrl{1} & \qw & \qw 
            \\ 
            & \gate{W_2(\bm{\theta}^2)}
            & \gate{R_{\mathrm{Y}}(0.1)}
            & \gate{R_{\mathrm{Y}}(0.2)}
            & \gate{R_{\mathrm{Y}}(0.4)}
            & \gate{W_1(\bm{\theta}^1)}
            & \qw
        }}
    }
    \caption{The circuit construction for \Cref{ex:3qubits}. (a) The given PQC $U$ with direct encoding of $x$ as a $\mathrm{Y}$-rotational gate. (b) The corresponding CPQC implementing the eight possible values of $\mathcal{X}$ depicted in the table as controlled $\mathrm{Y}$ rotations.}
    \label{fig:example_CVQC}
\end{figure}
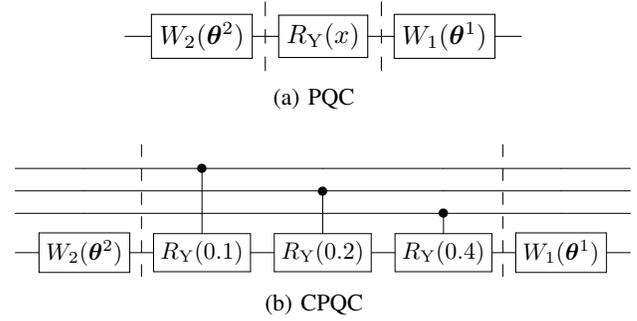
\end{example}

\begin{example}\label{efficient_cpqc_construction}
    \ \newline
    The previous example can be generalized to arbitrarily fine discretizations.
    Let $n \in \mathbb{N}$ be an arbitrary number of qubits used for the control register.
    Let our data $\mathcal{X} = \{x_1, \ldots, x_{2^{n-1}} \}$ be given by the
    discretization $x_k =  \tfrac{2 \pi}{2^{n}} k$. Each $x_k$ is encoded
    in the control register by the bit-representation of the integer $k$.
    Our PQC, $U(x,\bm{\theta}) = \prod_{l=1}^{L} S_l(x) W_l(\bm{\theta}^l)$, 
    can have arbitrary many layers, with each layer having an encoding circuit of the form
    $S_l(x) = R_{\alpha}(x), \alpha \in \{\mathrm{X}, \mathrm{Y}, \mathrm{Z}\}$. The parameterzied blocks $W_l(\bm{\theta}^l)$ can be chosen
    arbitrarily. In this setup, each encoding circuit $S_l(x)$ can be
    substituted by a control circuit $\Lambda_l(n)$ acting on $n+1$ qubits,
    \begin{align*}
        \Lambda_l(n) = \prod_{c=0}^{n-1}\Lambda(c,R_{\alpha}),
    \end{align*}
    such that the requirements of the \Cref{prop} are satisfied.
    Note that the construction of $\Lambda_l(n)$ uses $n$ factors; each factor
    can be constructed by two single qubit operations and two CNOTs.
    Thus, this construction scales like $O(n)$ and is efficient in the sense of \Cref{def:effciency}.
\end{example}

\section{Implementation of Genetic PQCs} \label{sec:implementation_of_genetic_pqcs}
VQAs show a lot of potential for enabling noisy intermediate-scale quantum devices to solve practical tasks.
As discussed in recent literature, choosing a suitable PQC is not a trivial task and has a huge influence on the performance of VQAs~\cite{duQuantumCircuitArchitecture2022}.
There are two important requirements while choosing the ansatz; it needs to be expressive enough to model the problem at hand and it needs to be as shallow as possible.
Unfortunately, these two goals often contradict.

An idea, that recently came up in different variations, is algorithmically adapting the ansatz to the problem at hand.
Some of the developed algorithms are ADAPT-VQE~\cite{grimsleyAdaptiveVariationalAlgorithm2019}, Rotoselect and Rotosolve~\cite{ostaszewskiStructureOptimizationParameterized2021}, Quantum Circuit Evolution of Augmenting Topologies (QCEAT)~\cite{huangRobustResourceefficientQuantum2022} and the already mentioned Variable Ansatzes (VAns)~\cite{bilkisSemiagnosticAnsatzVariable2023}.
Especially under the term Quantum Architecture Search (QAS) some recent research has been published~\cite{kuo2021quantum,Zhang_2022,duQuantumCircuitArchitecture2022}.
These works mainly focus on the VQE.
This means that the goal is to prepare one specific state, while we on the contrary want to learn functions.
The additional difficulty is integrating the dependency on the input data, also called the encoding, into an adaptive scheme.

In this section we will extend on the ideas of VAns.

\subsection{Variable Ansatz Algorithm}

The general idea is to iteratively alternate between adding randomly selected parameterized or encoding blocks, as defined in \cref{eq:blocks}, and removing noncontributing gates.

More formally the algorithm consists of:
\begin{itemize}
    \item Training data consisting of inputs \(\mathcal{X} = \{x_1, \dots, x_J\}\) and corresponding true values \(\mathcal{Y} = \{y_1, \dots, y_J\}\). Usually the \(y_i\) are real numbers and the \(x_i\) are either real values or real vectors.
    \item An initial parameterized circuit \(U_0(x, \theta_0)\), that depends on the input \(x\) and a parameter vector \(\theta_0\) of suitable size.
    \item A cost function \(C(\mathcal{X}, \mathcal{Y}, U, \theta)\) that gives some measure of prediction error. For example, the mean squared error between the prediction and the true values.
    \item A classical optimizer \(\Upsilon\) that minimizes the costs. Ideally \(\Upsilon(\mathcal{X}, \mathcal{Y}, U) = \min_\theta C(\mathcal{X}, \mathcal{Y}, U, \theta)\).
    \item A set of blocks that we can add to the circuit.
    \item A decision function \(g(c, \hat{c})\) that gives an acceptance probability given the old and the new costs.
    \item A reduction function \(\rho(U)\) that removes noncontributing gates.
    \item The number of iterations \(n_i\).
\end{itemize}

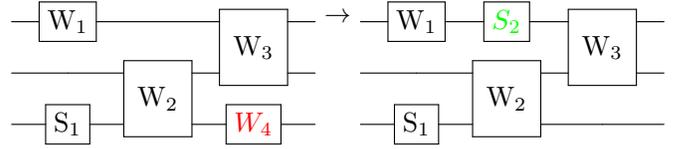
\begin{figure}[!t]
    \centering
    \begin{subfigure}{.47\linewidth}
        \centering
        \resizebox{\linewidth}{!}{
        \Qcircuit @C=1em @R=.7em {
                & \gate{\mathrm{W_1}} & \qw & \multigate{1}{\mathrm{W_3}} & \qw \\
                & \qw & \multigate{1}{\mathrm{W_2}} & \ghost{\mathrm{W_3}} & \qw\\
                & \gate{\mathrm{S_1}} & \ghost{\mathrm{W_2}} & \gate{\textcolor{red}{W_4}} & \qw\\
            }
        }
    \end{subfigure}
    {$\rightarrow$}%
    \begin{subfigure}{.47\linewidth}
        \centering
        \resizebox{\linewidth}{!}{
            \Qcircuit @C=1em @R=.7em {
                    & \gate{\mathrm{W_1}} & \gate{\textcolor{green}{S_2}} & \multigate{1}{\mathrm{W_3}} & \qw \\
                    & \qw & \multigate{1}{\mathrm{W_2}} & \ghost{\mathrm{W_3}} & \qw\\
                    & \gate{\mathrm{S_1}} & \ghost{\mathrm{W_2}} & \qw & \qw\\
                }
        }
    \end{subfigure}
    \caption{Schematic visualization of one step of \Cref{alg:structure_learning}, the block \(\mathrm{S_2}\) gets added while the block \(\mathrm{W_4}\) got removed.}
    \label{fig:structure_learning}
\end{figure}

\begin{algorithm}[H]
    \caption{Structure Learning}
    \label{alg:structure_learning}
    \begin{algorithmic}
        \STATE \(c_0 = \Upsilon(\mathcal{X}, \mathcal{Y}, U_0)\).
        \FOR{\(i = 1,\ldots, n_i\)}
            \STATE \textbf{Sample} block to add, qubits on which it operates, position in the circuit.
            \STATE \textbf{Construct} new circuit \(\hat{U}_i\).
            \STATE \textbf{Set} \(\hat{c}_i = \Upsilon(\mathcal{X}, \mathcal{Y}, \hat{U}_i)\).
            \STATE \textbf{Sample} Z uniformly distributed between \(0\) and \(1\).
            \IF{\(Z \le g(c_{i-1}, \hat{c}_i)\)}
                \STATE \(U_i = \rho(\hat{U}_i)\).
                \STATE \(c_i = \Upsilon(\mathcal{X}, \mathcal{Y}, U_i)\).
            \ELSE
                \STATE \(U_i = U_{i-1}\).
                \STATE \(c_i = c_{i-1}\).
            \ENDIF
        \ENDFOR
    \end{algorithmic}
\end{algorithm}
A qualitative demonstration of how the circuit changes during one step of this algorithm is given in \Cref{fig:structure_learning}.

In the initial proposal for that algorithm~\cite{bilkisSemiagnosticAnsatzVariable2023}, the set of possible blocks only consisted of parameterized unitaries \(U(\theta)\), for which there exists a \(\theta^*\) such that \(U(\theta^*)\) is the identity.
This ensures that adding these blocks to the circuit, while adding \(\theta^*\) to the parameter vector, does not change the circuit and, thus, does not increase the cost.
If then gradient descent is used to optimize the new set of parameters, there is a chance for a reduction of cost.
In conclusion, the cost is, not necessarily strictly, decreasing in each step.

For the parameter optimization of one PQC we use gradient based optimizers from the pennylane framework~\cite{bergholmPennyLaneAutomaticDifferentiation2022}. For the results of this work the gradients where calculated by automatic differentiation, which implies the use of noiseless simulators.

This algorithm allows for some modifications that will benefit the special requirements of our use case.
There are three main points we can modify.

\textbf{Blocks to add:}
While the original paper only focused on parameterized blocks that are used for training, we also allow encoding blocks.
They violate the requirement to be the identity with the right parameters, but will greatly improve the capability of approximating arbitrary functions. 
For datasets with more than one input feature, the set of blocks to sample from contains individual encoding blocks for every input feature.
This can also be interpreted as a form of feature selection.

\textbf{Gate Removal:}
By allowing a very slight increase in the cost, we can often leave out many gates.
As NISQ devices produce some error themselves, a slight increase in prediction error can be a good tradeoff for a simpler circuit.
By further preventing some of the gates to be removed during the iterations, we try to keep the search space as big as possible.
Examples for gates we do not remove are the last entanglement a qubit has with a measured one or the last parameterized gate on a qubit.

\textbf{Hardware Awareness:}
This approach also allows us to take the hardware into account, which the circuit should be executed on.
For example, we can only allow blocks of native gates and only entangle qubits that have a physical connection.
Additionally we can adapt the algorithm to limitations of NISQ devices by modifying the cost function to also take the depth of the circuit and the number of CNOT gates into account.

\subsection{Genetic Optimization}
In this section we propose to adapt techniques from classical genetic optimization, in order to improve the optimization of the circuit structure laid out in \Cref{alg:structure_learning}.

Genetic optimization keeps track of a population of possible solutions and applies three concepts on this population~\cite{melanieIntroductionGeneticAlgorithms}:
\begin{itemize}
    \item Mutation: Randomly modify one possible solution.
    \item Crossover: Combine different solutions to form a new one.
    \item Selection: Deciding which solutions will be used in the next generation.
\end{itemize}
In our approach, the crossover step is problematic, as it is not clear if and why the combination of two parameterized circuits should perform better than the individual ones.
Thus, we focus on mutation and selection.
In addition to the inputs from \Cref{alg:structure_learning}, we need a number of generations \(n_g\) and the size of the population \(n_p\).
\begin{algorithm}[H]
    \caption{Genetic Structure Learning}
    \label{alg:genetic_structure_learning}
    \begin{algorithmic}
        \STATE Initialize circuits \(\mathbf{U}_0 = (U_0^j)_{j=1,\dots,n_p}\).
        \FOR{i = 1,\ldots, \(n_g\)}
            \STATE \textbf{Mutation}: Apply \Cref{alg:structure_learning} to all elements of \(\mathbf{U}_{i-1}\) and store cost of each circuit \(c_{i-1}\).
            \STATE \textbf{Selection}: Sample new generation \(\mathbf{U}_i\) by sampling \(n_p\) times from \(\mathbf{U}_{i-1}\) where the probability to choose \(U_{i-1}^j\) is given by \(w_j = \frac{1}{c_j^2 \sum_{k=1}^{n_p} \frac{1}{c_k^2}}\).
        \ENDFOR
        \STATE Choose the best circuit from \(\mathbf{U}_{n_g}\) as output.
    \end{algorithmic}
\end{algorithm}

\subsection{Structure Learning for CPQCs}
For the application part we want to take circuits generated by \Cref{alg:structure_learning,,alg:genetic_structure_learning} and build controlled versions of them.
Here we will demonstrate which choices we make in these algorithms to be able to do just that.
At first, we require that the initial circuit is of the form \cref{eq:pqc}. As this form is very general, this is not a real restriction.
Further, we only allow blocks of the structure \cref{eq:blocks} to be added.
In every step we only add one encoding or parameterized block, but this can still be treated as one layer, as the identity is a valid example for both kinds of blocks.
We restrict the encoding blocks to simple rotations, i.e., the \(T_{l,k}\) 
are chosen as the identity and \(H_{l,k}\) is as one of \(\sigma_x/2\), \(\sigma_y/2\) or \(\sigma_z/2\).
At last, we have to make sure to not break the structure by removing individual gates in the reduction step.
By increasing the number of layers, we can write every individual gate as one layer.
Thus, removing a single gate, i.e., one layer, we keep the structure from \cref{eq:pqc}.
This enables us to use the encoding scheme described in \Cref{sec:cpqc}.

\section{Application to Option Pricing}\label{sec:application_to_options}

We apply the presented methodology to the problem of derivative pricing in finance and compare our method to current state-of-the-art quantum algorithms. Assume that in general we have underlying securities $X$ with some distribution $\mathcal{P}$ under a fitting probability measure $\mathbb{Q}$, and an option payoff $f(X)$. Assume $x_k \in [0,2\pi)$ is a scaled discretization of $X$ with $k\in \{0,\dots,2^n-1\}$, and $b(x_k)$ its bit representation. 
Recall, that we assume the operator $P$ implementing the distribution to be given. We present an efficient implementation of the operator $F$ to construct the needed quantum state (c.f., \eqref{eq:FP_application_state})

\begin{align*}
    \ket{\varphi_0}_c \ket{0}_t
    +
    \sum_{k=0}^{2^n-1} \sqrt{p_k} \ket{b(x_k)}_c \sqrt{f(x_k)} \ket{1}_t,
\end{align*}
where $\ket{\varphi_0}$ is an arbitrary state.

\subsection{European Vanilla Options}\label{subsec:european_vanilla}

\begin{figure}[!t]
    \centering
    \includegraphics[width=\linewidth]{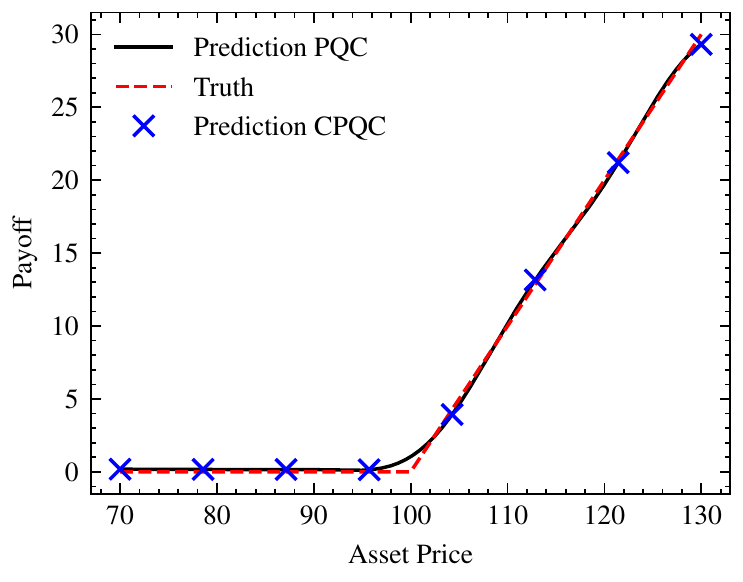}
    \caption{Function approximation by a PQC that was found by \Cref{alg:genetic_structure_learning} and its CPQC version. The function approximated is the payoff of an European Call with a Strike of 100.}
    \label{fig:european_call_prediction}
\end{figure}

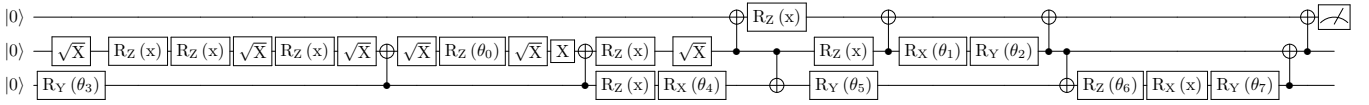
\begin{figure*}[!t]
    \centering
    \resizebox{\linewidth}{!}{
    \Qcircuit @C=0.2em @R=0.3em @!R { \\
	 	\nghost{\ket{{0}} } & \lstick{\ket{{0}} } & \qw & \qw & \qw & \qw & \qw & \qw & \qw & \qw & \qw & \qw & \qw & \qw & \qw & \qw & \targ & \gate{\mathrm{R_Z}\,(\mathrm{x})} & \qw & \targ & \qw & \qw & \targ & \qw & \qw & \qw & \qw & \qw & \targ & \meter \\
	 	\nghost{\ket{{0}} } & \lstick{\ket{{0}} } & \gate{\mathrm{\sqrt{X}}} & \gate{\mathrm{R_Z}\,(\mathrm{x})} & \gate{\mathrm{R_Z}\,(\mathrm{x})} & \gate{\mathrm{\sqrt{X}}} & \gate{\mathrm{R_Z}\,(\mathrm{x})} & \gate{\mathrm{\sqrt{X}}} & \targ & \gate{\mathrm{\sqrt{X}}} & \gate{\mathrm{R_Z}\,(\mathrm{\theta_0})} & \gate{\mathrm{\sqrt{X}}} & \gate{\mathrm{X}} & \targ & \gate{\mathrm{R_Z}\,(\mathrm{x})} & \gate{\mathrm{\sqrt{X}}} & \ctrl{-1} & \ctrl{1} & \gate{\mathrm{R_Z}\,(\mathrm{x})} & \ctrl{-1} & \gate{\mathrm{R_X}\,(\mathrm{\theta_1})} & \gate{\mathrm{R_Y}\,(\mathrm{\theta_2})} & \ctrl{-1} & \ctrl{1} & \qw & \qw & \qw & \targ & \ctrl{-1} & \qw \\
	 	\nghost{\ket{{0}} } & \lstick{\ket{{0}} } & \gate{\mathrm{R_Y}\,(\mathrm{\theta_3})} & \qw & \qw & \qw & \qw & \qw & \ctrl{-1} & \qw & \qw & \qw & \qw & \ctrl{-1} & \gate{\mathrm{R_Z}\,(\mathrm{x})} & \gate{\mathrm{R_X}\,(\mathrm{\theta_4})} & \qw & \targ & \gate{\mathrm{R_Y}\,(\mathrm{\theta_5})} & \qw & \qw & \qw & \qw & \targ & \gate{\mathrm{R_Z}\,(\mathrm{\theta_6})} & \gate{\mathrm{R_X}\,(\mathrm{x})} & \gate{\mathrm{R_Y}\,(\mathrm{\theta_7})} & \ctrl{-1} & \qw & \qw \\
\\ }}
    \caption{The trained PQC for the payoff of a European call option generated by \Cref{alg:genetic_structure_learning}. \(\bm{\theta}\) = [0.93963, 2.51976, -0.30702, -0.22985, -0.302, -0.09293, 0.15291, 0.05979]}
    \label{fig:european_call_circuit}
\end{figure*}

When pricing European vanilla options, we have a single underlying security $X = X^{(1)}$ with a fixed maturity $T$ we are omitting in the notation for simplicity. Then, the payoff of a call option $f_+$ or put option $f_-$ with a fixed strike $K$ is defined as
\begin{align*}
    f_{\pm}(X) \coloneqq \max \left( \pm(X - K), 0 \right).
\end{align*}
The option price can be calculated by discounting the expected payoff \cite{korn_monte_2010}. We assume this discounting to be constant and applied in post-processing on a classical device. We denote with capital $X_k, k=0,\dots,2^n-1$ the discretized version of $X$.

\textbf{Quantum arithmetic:}
We compare our method to the implementation by IBM's qiskit finance module \cite{Qiskit}, based on the current state-of-the-art method introduced by References \cite{Woerner_2019,Stamatopoulos2020optionpricingusing}. They priced a vanilla option by using a quantum comparator circuit that flips an ancilla qubit initially in state $\ket{0}$ to the state $\ket{1}$ if $X_i > K$, using $O(n)$ gates.
Afterwards, introducing a scaling factor $\tilde{c}$ they use the approximation $\sin^2(y+\pi/4) \approx y + 1/2$ for small $|y|$ to efficiently implement the $\pm (X - K)$ by linearly many Y-rotational gates, each controlled by only two qubits, on a second ancilla qubit.
For the European call option, the resulting probability to measure this second ancilla qubit in state $\ket{1}$ is given by
\begin{align*}
    \frac{1}{2} - \tilde{c} + \frac{2\tilde{c}}{\max_i X_i - K} \sum_{X_i \geq K} p_i (X_i - K),
\end{align*}
where the option price can be recovered by rescaling. The entire quantum arithmetic implementation scales as $O(n)$ in the qubit register size $n$, i.e., in $O(\log(N))$ for $N=2^n$ discretization points of the underlying.

\begin{figure}[t]
    \centering
    \includegraphics[width=\linewidth]{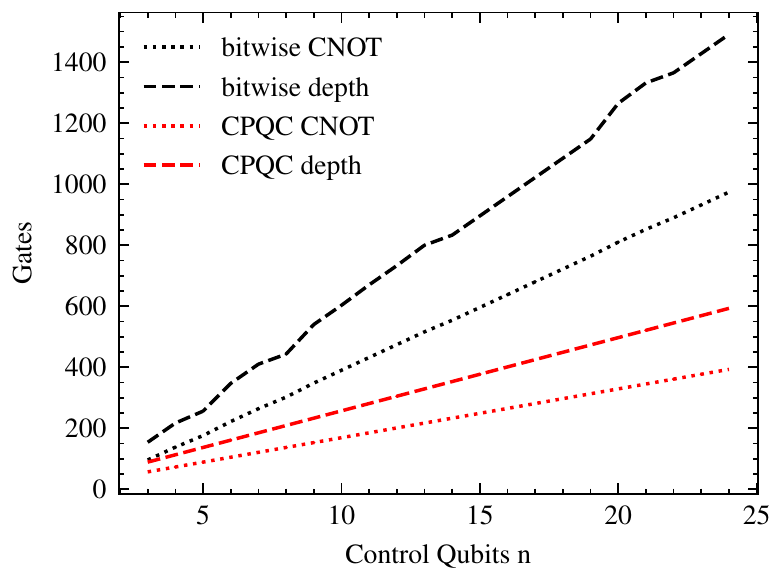}
    \caption{Total CNOT count and depth comparison of the quantum arithmetic approach by IBM and our CPQC method explained in \Cref{subsec:european_vanilla}. Counts were computed with respect to the CNOT and $\mathrm{U}(\theta, \varphi, \lambda)$ gate decomposition.}
    \label{fig:benchmark_call_option}
\end{figure}

\textbf{CPQC:}
We used \Cref{alg:genetic_structure_learning} to generate the circuit seen in \Cref{fig:european_call_circuit}, with the corresponding CPQC being shown in \Cref{fig:CPQC_european_call}.
The resulting function approximation is demonstrated in \Cref{fig:european_call_prediction}.
Since a PQC can be represented as a truncated Fourier series, where the frequencies included depend on the extent of data reuploading~\cite{schuld2021effect}, we observe a smoothing of the learned function at the one non differentiable point.
The genetic algorithm simulated \(20\) generations with a population size of \(48\).
Each generation did \(20\) iterations of structure optimization.
The parameter optimization was done by Pennylanes Root mean squared propagation optimizer.
A new step was accepted if the costs decreased; whereas if the costs increased the probability to accept the new structure was calculated as \(\exp(-5 (c_{new} - c_{old})/c_{old})\).
A gate was removed if the removal lead to an increase in cost below 1\%.
The whole procedure took about \(5.5\) hours on a dual Intel Xeon Gold 6240R system with 48 cores.
We can see a comparison of the required number of CNOT gates and depth scaling of both decomposed quantum circuits in \Cref{fig:benchmark_call_option}. 
The decomposition is done with regards to CNOT gates and $\mathrm{U}(\theta, \varphi, \lambda)$ gates.

\subsection{Basket Option - Fixed Weight}\label{subsec:basked_fixed_weight}

\begin{figure}[t]
    \centering
    \includegraphics[width=\linewidth]{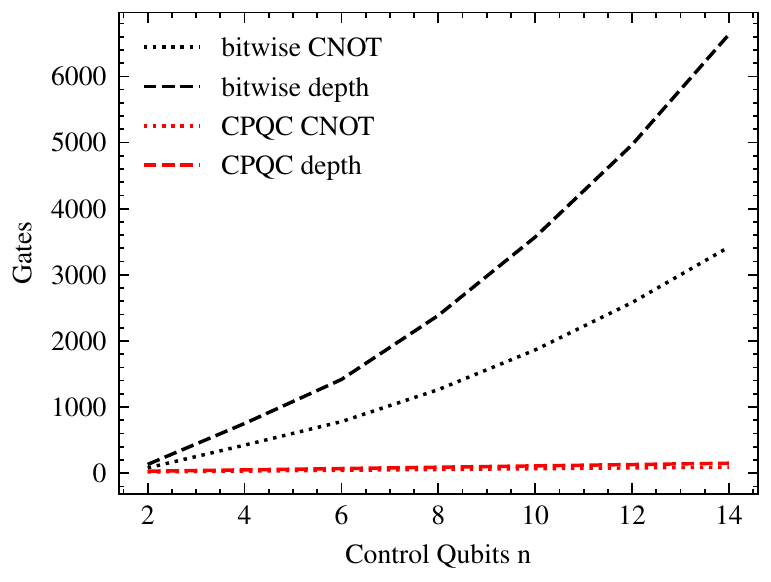}
    \caption{Total CNOT count and depth comparison of the quantum arithmetic approach by IBM and our CPQC method explained in \Cref{subsec:basked_fixed_weight}. We chose two underlying securities $\bm{X}=(X^{(1)}, X^{(2)})$, so $\mathcal{K}=2$ and $n_k=n/2$. The weight was set to $\bm{w}=(2/3,1/3)$, which was implemented for the quantum arithmetic algorithm for the scaled up integer values $w_1=2,w_2=1$. Counts were computed with respect to the CNOT and $\mathrm{U}(\theta, \varphi, \lambda)$ gate decomposition. For $n=14$, the CPQC has a CNOT count of $92$ and a depth of $151$.}
    \label{fig:benchmark_basket_option}
\end{figure}

A basket option is defined for a vector of underlying securities $\bm{X} = (X^{(1)},...,X^{(\mathcal{K})})^T$, fixed weights $\pmb{w}=(w_1,\dots,w_{\mathcal{K}}), w_k\in[0,1], \sum_k w_k = 1$, and a fixed strike $K$. Its payoff profile is given by
\begin{align*}
    f_\text{Basket}(\bm{X}) 
    \coloneqq
    \max\left( \bm{w} \bm{X} - K, 0 \right).
\end{align*}

\textbf{Quantum arithmetic:}
Again, we compare our approach to the implementation by IBM qiskit based on Reference \cite{Stamatopoulos2020optionpricingusing}. Each underlying $X^{(i)}$ is represented by its discretization on a separate qubit register $r_i$, in total needing $n$ qubits.
The circuit to implement the payoff is now similar to the vanilla option, except that an additional operator computing the weighted sum of the qubit registers is needed before applying the comparator. This sum is saved into a new set of ancilla qubits of size $\log_2 (\sum_k w_k) + 1$.
Then the methodology of the vanilla option is applied to this new register containing the weighted sum.
Because of this weighted sum operator, the discretization for all securities has to be the same. It needs $O(n \log_2 n)$ gates, which is also the total complexity of the quantum arithmetic implementation of $F$.

\textbf{CPQC:}
This task shows the main improvement of the CPQC approach. For a chosen level of approximation, the PQC has a fixed qubit and gate number. 
Coupling the PQC with the control register via the method described in \Cref{efficient_cpqc_construction} only adds a linear amount of gates, resulting in a total gate complexity for the CPQC of $O(n)$. The genetically trained PQC and corresponding CPQC for this task are presented in \Cref{fig:european_basket_call_circuit} and \Cref{fig:european_basket_call_circuit_controlled}, respectively. The PQC approximation, as well as the $f_\text{Basket}$, can be found in \Cref{fig:basket_call_fixed_weight_functions}.

With the same procedure used as for the call option, we can see a comparison of the required number of CNOT gates and depth scaling of both decomposed quantum circuits in \Cref{fig:benchmark_basket_option}.

\subsection{Basket Option - Variable Weight}

\begin{figure}[!t]
    \centering
    \includegraphics[width=\linewidth]{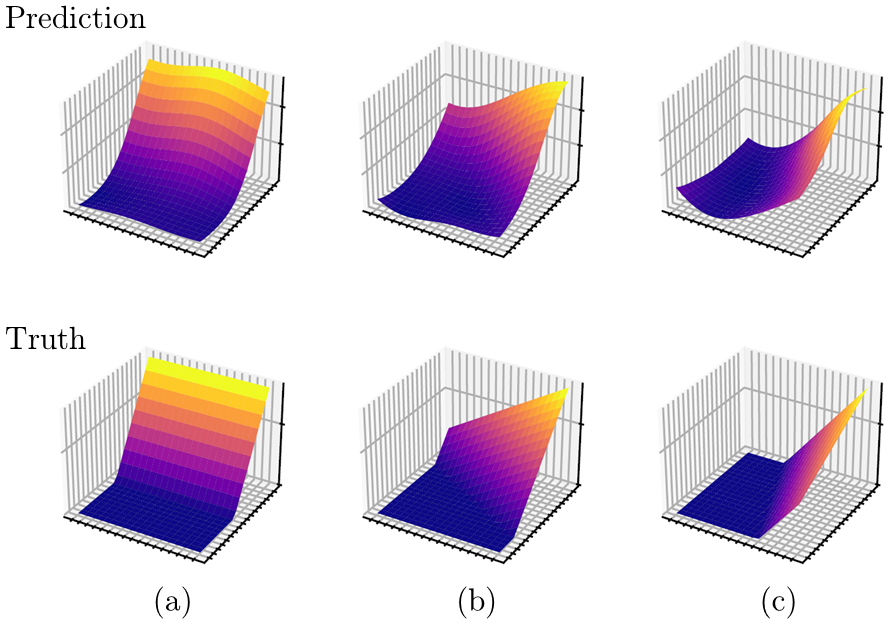}
    \caption{Result of the circuit from \Cref{fig:european_basket_call_variable_weight_circuit} in the upper row compared with the true payoff function $f_\text{Basket}$ in the lower row. This circuit takes three inputs, the value of the two assets and the weight of the first asset. The figure shows the results three different weights: (a) $w_1=0$, (b) $w_1\approx 0.37$, and (c) $w_1\approx 0.74$.}
    \label{fig:variable_weight_basket_option}
\end{figure}

Although it substantially increases the difficulty for the quantum arithmetic implementation, it is straightforward to include a variable weight for the basket option as an input feature of our PQC, hence, of the CPQC. 
By doing so, one could pre-train a sophisticated PQC for a specific set of underlyings and a corresponding resolution. 
This, albeit possibly needing a larger circuit, has a resulting CPQC that has a gate scaling linear in the number of state qubits needed to represent the securities $X^{(i)}$.
This CPQC can then either be used for a variety of different weights (in a range that was previously used for training), or for computing the prices of a basket where the weights depend on some different state register.
The result of our genetically trained PQC, depicted in \Cref{fig:european_basket_call_variable_weight_circuit}, can be seen in \Cref{fig:variable_weight_basket_option}. It is straightforward to construct the CPQC from this PQC as all the encoding blocks are rotational gates.
Of course, the prediction could be improved by employing a larger PQC at the cost of more depth.
At least for two underlyings, it should be emphasized that the variable weight $w_1$ is only encoded once in the PQC. It seems to rotate the payoff function, which the PQC seems to be able to implement well.
These observations might also be useful for finding circuits, that implement different payoffs, \textit{by hand}. 

Additionally, this implementation of a basket option might open up the possibility for pricing index options that have a combination of stocks as the underlying. We leave this for future research.

\section{Conclusion}\label{sec:conclusion}
We presented an automated approach to the function implementation for the quantum alternative of classical Monte Carlo algorithms based on quantum machine learning.
First, laying the theoretical groundwork to build a conditional variant, we introduced our own genetic algorithm to find a suitable PQC.
Second, we compared our approach to current quantum algorithms for the task of option pricing, which rely on a quantum version of bitwise computations.
We were able to show that at the cost of a classical pre-training and a preemptively chosen approximation level, we generated reusable CPQCs that need a fraction of the quantum resources compared to the bitwise algorithms.

Possible future research is the further improvement of the genetic optimization of the PQC.
One possible direction is taking the hardware into account by using only native gates, only entangling physically connected qubits, weighting approximation error against hardware noise and also optimizing for robustness in the presence of noise.
Additionally one could adapt the algorithm to make sure that the inverse of the function operator $F$ is also efficient, which would be a benefit for the application of AE.
In the same vein, it should be researched if the application of PQCs translates to shortcuts one could take for quantum Monte Carlo integration algorithms. This has previously been done for other approaches, e.g., spin-echo circuit optimization \cite{vazquez_efficient_2020}.
The CPQC approach might also make existing AE techniques that have additional conditions on their function operator easier to implement \cite{Plekhanov_2022}.
Of course, a sophisticated ressource estimation for quantum advantage including the CPQC approach should be made as well, c.f., \cite{chakrabarti_threshold_2021}.

We are currently investigating the possibility of generalizing the conditional variant of PQCs for encodings other than simple rotations, which will probably translate to a slight encoding error, and more sophisticated control circuits $\Lambda$.
Nevertheless, in the current NISQ environment small approximation errors perish in comparison to noise introduced by quantum gates. 
It has to be studied at which level of reduction in gate counts this additional error is justified.


\bibliographystyle{IEEEtran}
\bibliography{bibliography}

\renewcommand{\thesection}{A.{section}}
\renewcommand\thefigure{A.\arabic{figure}}
\setcounter{figure}{0}
\section*{Appendix}
\begin{figure}[h]
    \centering
    \includegraphics[width=\linewidth]{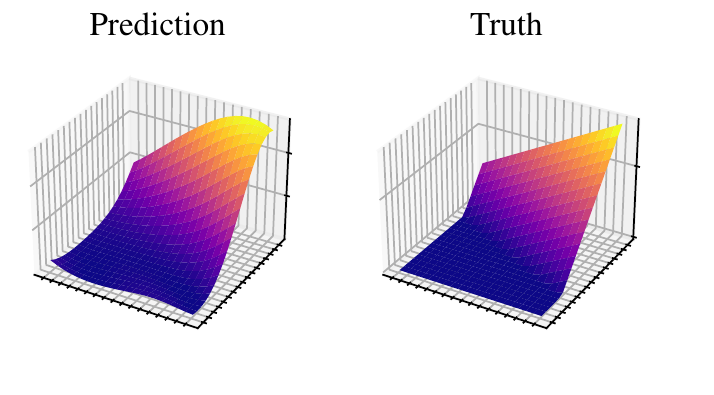}
    \caption{Payoff of a European basket call. On the left, the approximation by the circuit from \Cref{fig:european_basket_call_circuit}, on the right the true payout. On the X and Y axis are the prices of the two assets and on the Z axis the payoff.}
    \label{fig:basket_call_fixed_weight_functions}
\end{figure}

\begin{proof_prop}\label{prop_appendix}
    \ \newline
    For the sake of brevity, we write
    $\ket{b(\bm{x})}_c := \bigotimes_{k} \ket{b(\bm{x}_k)}_{r_k}$.
    First, we note that the assumption implies that $U$ and $C_n(U)$
    will generate the same quantum state when some data $\bm{x} \in \mathcal{X}$
    is input,
    \begin{align}\label{eq:lemma}
        (I_n \otimes U)\ket{b(\bm{x})}_c \ket{0}_t
        = C_n(U) \ket{b(\bm{x})}_c \ket{0}_t.
    \end{align}
    Then, for the right hand side of \cref{eq:cpqc_goal}, it follows that
    \begin{align*}
    & f_{C_n(U),P}(\bm{\theta}) \\
    & = \sum_{i,j=0}^{2^n-1}\sqrt{p_i}\sqrt{p_j}
    \bra{b(x_i)}_c \bra{0}_t C_n(U)^{\dagger}
    \mathcal{M}
    C_n(U)\ket{b(x_j)}_c\ket{0}_t \\
    & \stackrel{\cref{eq:lemma}}{=} \sum_{i,j=0}^{2^n-1}\sqrt{p_i}\sqrt{p_j}
    \bra{b(x_i)}_c \bra{0}_t (I_n \otimes U)^{\dagger}
    \mathcal{M}
    (I_n \otimes U)\ket{b(x_j)}_c\ket{0}_t \\
    & = \sum_{i=0}^{2^n-1} p_i \bra{0}_t U^{\dagger}M U\ket{0}_t\\
    & = \sum_{i=0}^{2^n-1} p_i f_U(\bm{x}^i,\bm{\theta}). 
    \end{align*}
    \hfill $\square$
\end{proof_prop}

\begin{landscape}
\begin{figure}
    \centering
    \resizebox{\linewidth}{!}{
        \Qcircuit @C=0.2em @R=0.3em @!R { \\
        \nghost{\ket{{0}} } & \lstick{\ket{{0}} } & \multigate{2}{P_1} & \ctrl{4} & \qw & \qw & \ctrl{4} & \qw & \qw & \qw & \ctrl{4} & \qw & \qw & \qw & \qw & \qw & \qw & \qw & \qw & \qw & \ctrl{4} & \qw & \ctrl{5} & \qw & \qw & \qw & \qw & \qw & \ctrl{3} & \qw & \ctrl{4} & \qw & \qw & \qw & \qw & \qw & \qw & \qw & \qw & \qw & \ctrl{5} & \qw &  \qw & \qw & \qw & \qw & \qw & \qw & \qw & \qw\\
        \nghost{\ket{{0}} } & \lstick{\ket{{0}} } & \ghost{P_1} & \qw & \ctrl{3} & \qw & \qw & \ctrl{3} & \qw & \qw & \qw & \ctrl{3} & \qw & \qw & \qw & \qw & \qw & \qw & \qw & \qw & \qw & \ctrl{3} & \qw & \qw & \ctrl{4} & \qw & \qw & \qw & \qw & \ctrl{2} & \qw & \qw & \ctrl{3} & \qw & \qw & \qw & \qw & \qw & \qw & \qw & \qw & \ctrl{4} & \qw & \qw &  \qw & \qw & \qw & \qw & \qw & \qw\\
        \nghost{\ket{{0}} } & \lstick{\ket{{0}} } & \ghost{P_1} & \qw & \qw & \ctrl{2} & \qw & \qw & \ctrl{2} & \qw & \qw & \qw & \ctrl{2} & \qw & \qw & \qw & \qw & \qw & \qw & \qw & \qw & \qw & \qw & \ctrl{2} & \qw & \ctrl{3} & \qw & \qw & \qw & \qw & \qw & \ctrl{1} & \qw & \ctrl{2} & \qw & \qw & \qw & \qw & \qw & \qw & \qw & \qw & \qw & \ctrl{3} & \qw & \qw &  \qw & \qw & \qw & \qw\\
        \nghost{\ket{{0}} } & \lstick{\ket{{0}} } & \qw & \qw & \qw & \qw & \qw & \qw & \qw & \qw & \qw & \qw & \qw & \qw & \qw & \qw & \qw & \qw & \qw & \qw & \qw & \qw & \qw & \qw & \qw & \qw & \qw & \targ & \gate{\mathrm{R_Z}\,(\mathrm{\frac{4\pi}{7}})} & \gate{\mathrm{R_Z}\,(\mathrm{\frac{2\pi}{7}})} & \qw & \gate{\mathrm{R_Z}\,(\mathrm{\frac{\pi}{7}})} & \qw & \qw & \targ & \qw & \qw & \targ & \qw & \qw & \qw & \qw & \qw & \qw & \qw & \qw & \qw & \qw & \targ & \meter \\
        \nghost{\ket{{0}} } & \lstick{\ket{{0}} } & \gate{\mathrm{\sqrt{X}}} & \gate{\mathrm{R_Z}\,(\mathrm{\frac{4\pi}{7}})} & \gate{\mathrm{R_Z}\,(\mathrm{\frac{2\pi}{7}})} & \gate{\mathrm{R_Z}\,(\mathrm{\frac{\pi}{7}})} & \gate{\mathrm{R_Z}\,(\mathrm{\frac{4\pi}{7}})} & \gate{\mathrm{R_Z}\,(\mathrm{\frac{2\pi}{7}})} & \gate{\mathrm{R_Z}\,(\mathrm{\frac{\pi}{7}})} & \gate{\mathrm{\sqrt{X}}} & \gate{\mathrm{R_Z}\,(\mathrm{\frac{4\pi}{7}})} & \gate{\mathrm{R_Z}\,(\mathrm{\frac{2\pi}{7}})} & \gate{\mathrm{R_Z}\,(\mathrm{\frac{\pi}{7}})} & \gate{\mathrm{\sqrt{X}}} & \targ & \gate{\mathrm{\sqrt{X}}} & \gate{\mathrm{R_Z}\,(\mathrm{\theta_0})} & \gate{\mathrm{\sqrt{X}}} & \gate{\mathrm{X}} & \targ & \gate{\mathrm{R_Z}\,(\mathrm{\frac{4\pi}{7}})} & \gate{\mathrm{R_Z}\,(\mathrm{\frac{2\pi}{7}})} & \qw & \gate{\mathrm{R_Z}\,(\mathrm{\frac{\pi}{7}})} & \qw & \qw & \gate{\mathrm{\sqrt{X}}} & \ctrl{-1} & \ctrl{1} & \qw & \gate{\mathrm{R_Z}\,(\mathrm{\frac{4\pi}{7}})} & \qw & \gate{\mathrm{R_Z}\,(\mathrm{\frac{2\pi}{7}})} & \gate{\mathrm{R_Z}\,(\mathrm{\frac{\pi}{7}})} & \ctrl{-1} & \gate{\mathrm{R_X}\,(\mathrm{\theta_1})} & \gate{\mathrm{R_Y}\,(\mathrm{\theta_2})} & \ctrl{-1} & \ctrl{1} & \qw & \qw & \qw & \qw & \qw & \qw & \qw & \qw & \targ & \ctrl{-1} & \qw\\
        \nghost{\ket{{0}} } & \lstick{\ket{{0}} } & \gate{\mathrm{R_Y}\,(\mathrm{\theta_3})} & \qw & \qw & \qw & \qw & \qw & \qw & \qw & \qw & \qw & \qw & \qw & \ctrl{-1} & \qw & \qw & \qw & \qw & \ctrl{-1} & \qw & \qw & \gate{\mathrm{R_Z}\,(\mathrm{\frac{4\pi}{7}})} & \qw & \gate{\mathrm{R_Z}\,(\mathrm{\frac{2\pi}{7}})} & \gate{\mathrm{R_Z}\,(\mathrm{\frac{\pi}{7}})} & \gate{\mathrm{R_X}\,(\mathrm{\theta_4})} & \qw & \targ & \gate{\mathrm{R_Y}\,(\mathrm{\theta_5})} & \qw & \qw & \qw & \qw & \qw & \qw & \qw & \qw & \targ & \gate{\mathrm{R_Z}\,(\mathrm{\theta_6})} & \gate{\mathrm{R_X}\,(\mathrm{\frac{4\pi}{7}})} & \gate{\mathrm{R_X}\,(\mathrm{\frac{2\pi}{7}})} & \qw & \gate{\mathrm{R_X}\,(\mathrm{\frac{\pi}{7}})} & \qw & \gate{\mathrm{R_Y}\,(\mathrm{\theta_7})} & \qw & \ctrl{-1} & \qw & \qw\\
    \\ }}
    \caption{Controlled version of \Cref{fig:european_call_circuit}. A three qubit register is used as control.}
    \label{fig:CPQC_european_call}

    \resizebox{\linewidth}{!}{
        \Qcircuit @C=0.2em @R=0.3em @!R { \\
        \nghost{\ket{{0}}} & \lstick{\ket{{0}}} & \qw & \qw & \qw & \qw & \qw & \qw & \qw & \qw & \qw & \qw & \qw & \qw & \qw & \qw & \qw & \qw & \qw & \qw & \targ & \meter \\
        \nghost{\ket{{0}}} & \lstick{\ket{{0}}} & \qw & \qw & \qw & \qw & \qw & \qw & \qw & \targ & \ctrl{2} & \gate{\mathrm{R_Z}\,(\mathrm{\theta_8})} & \ctrl{2} & \targ & \qw & \qw & \qw & \qw & \qw & \targ & \ctrl{-1} & \qw \\
        \nghost{\ket{{0}}} & \lstick{\ket{{0}}} & \gate{\mathrm{R_Y}\,(\mathrm{\theta_0})} & \gate{\mathrm{R_X}\,(\mathrm{x_0})} & \gate{\mathrm{R_Z}\,(\mathrm{\theta_1})} & \ctrl{1} & \gate{\mathrm{R_Y}\,(\mathrm{\theta_2})} & \gate{\mathrm{R_Z}\,(\mathrm{x_0})} & \gate{\mathrm{R_Z}\,(\mathrm{x_1})} & \ctrl{-1} & \qw & \gate{\mathrm{R_Y}\,(\mathrm{x_1})} & \qw & \ctrl{-1} & \gate{\mathrm{R_Z}\,(\mathrm{\theta_3})} & \gate{\mathrm{R_Y}\,(\mathrm{\theta_4})} & \qw & \qw & \targ & \ctrl{-1} & \qw & \qw\\
        \nghost{\ket{{0}} } & \lstick{\ket{{0}} } & \qw & \qw & \qw & \targ & \qw & \qw & \qw & \qw & \targ & \qw & \targ & \gate{\mathrm{R_Y}\,(\mathrm{\theta_5})} & \gate{\mathrm{R_Z}\,(\mathrm{x_0})} & \gate{\mathrm{R_Y}\,(\mathrm{\theta_6})} & \gate{\mathrm{R_Z}\,(\mathrm{x_1})} & \gate{\mathrm{R_Y}\,(\mathrm{\theta_7})} & \ctrl{-1} & \qw & \qw & \qw\\
    \\ }}
    \caption{The genetically trained PQC for the payoff of a European basket option with fixed weight generated by \Cref{alg:structure_learning}. \(\bm{\theta}\) = [1.70789, -1.71191, -1.11194, -0.72190, -0.74404, 1.40678, 0.67897, 0.17417, -0.16803]}
    \label{fig:european_basket_call_circuit}
\end{figure}

\begin{figure}
    \centering
    \resizebox{\linewidth}{!}{
        \Qcircuit @C=0.2em @R=0.3em @!R { \\
        \nghost{\ket{{0}} } & \lstick{\ket{{0}} } & \multigate{1}{P_1} & \ctrl{7} & \qw & \qw & \qw & \qw & \ctrl{7} & \qw & \qw & \qw & \qw & \qw & \qw & \qw & \qw & \qw & \qw & \qw & \qw & \qw & \ctrl{8} & \qw & \qw &  \qw & \qw & \qw & \qw & \qw & \qw & \qw & \qw & \qw & \qw & \qw & \qw & \qw & \qw & \qw & \qw\\
        \nghost{\ket{{0}} } & \lstick{\ket{{0}} } & \ghost{P_1} & \qw & \ctrl{6} & \qw & \qw & \qw & \qw & \ctrl{6} & \qw & \qw & \qw & \qw & \qw & \qw & \qw & \qw & \qw & \qw & \qw & \qw & \qw & \qw & \ctrl{7} & \qw & \qw &  \qw & \qw & \qw & \qw & \qw & \qw & \qw & \qw & \qw & \qw & \qw & \qw & \qw & \qw\\
        \nghost{\ket{{0}} } & \lstick{\ket{{0}} } & \multigate{2}{P_2} & \qw & \qw & \qw & \qw & \qw & \qw & \qw & \ctrl{5} & \qw & \qw & \qw & \qw & \ctrl{5} & \qw & \qw & \qw & \qw & \qw & \qw & \qw & \qw & \qw & \qw & \qw & \qw & \ctrl{6} & \qw &  \qw & \qw & \qw & \qw & \qw & \qw & \qw & \qw & \qw & \qw & \qw\\
        \nghost{\ket{{0}} } & \lstick{\ket{{0}} } & \ghost{P_2} & \qw & \qw & \qw & \qw & \qw & \qw & \qw & \qw & \ctrl{4} & \qw & \qw & \qw & \qw & \qw & \ctrl{4} & \qw & \qw & \qw & \qw & \qw & \qw & \qw & \qw & \qw & \qw & \qw & \ctrl{5} & \qw & \qw &  \qw & \qw & \qw & \qw & \qw & \qw & \qw & \qw & \qw\\
        \nghost{\ket{{0}} } & \lstick{\ket{{0}} } & \ghost{P_2} & \qw & \qw & \qw & \qw & \qw & \qw & \qw & \qw & \qw & \ctrl{3} & \qw & \qw & \qw & \qw & \qw & \ctrl{3} & \qw & \qw & \qw & \qw & \qw & \qw & \qw & \qw & \qw & \qw & \qw & \qw & \ctrl{4} & \qw & \qw &  \qw & \qw & \qw & \qw & \qw & \qw & \qw\\
        \nghost{\ket{{0}} } & \lstick{\ket{{0}} } & \gate{\mathrm{R_Y}\,(\mathrm{\theta_0})} & \qw & \qw & \qw & \qw & \qw & \qw & \qw & \qw & \qw & \qw & \qw & \qw & \qw & \qw & \qw & \qw & \qw & \qw & \qw & \qw & \qw & \qw & \qw & \qw & \qw & \qw & \qw & \qw & \qw & \qw & \qw & \qw & \qw & \qw & \targ &  \qw & \qw & \meter\\
        \nghost{\ket{{0}} } & \lstick{\ket{{0}} } & \qw & \qw & \qw & \qw & \qw & \qw & \qw & \qw & \qw & \qw & \qw & \targ & \ctrl{2} & \qw & \gate{\mathrm{R_Z}\,(\mathrm{\theta_9})} & \qw & \qw & \ctrl{2} & \targ & \qw & \qw & \qw & \qw & \qw & \qw & \qw & \qw & \qw & \qw & \qw & \qw & \qw & \qw & \qw & \targ & \ctrl{-1} & \qw & \qw & \qw\\
        \nghost{\ket{{0}} } & \lstick{\ket{{0}} } & \gate{\mathrm{R_Y}\,(\mathrm{\theta_1})} & \gate{\mathrm{R_X}\,(\mathrm{\frac{2\pi}{3}})} & \gate{\mathrm{R_X}\,(\mathrm{\frac{\pi}{3}})} & \gate{\mathrm{R_Z}\,(\mathrm{\theta_2})} & \ctrl{1} & \gate{\mathrm{R_Y}\,(\mathrm{\theta_3})} & \gate{\mathrm{R_Z}\,(\mathrm{\frac{2\pi}{3}})} & \gate{\mathrm{R_Z}\,(\mathrm{\frac{\pi}{3}})} & \gate{\mathrm{R_Z}\,(\mathrm{\frac{4\pi}{7}})} & \gate{\mathrm{R_Z}\,(\mathrm{\frac{2\pi}{7}})} & \gate{\mathrm{R_Z}\,(\mathrm{\frac{\pi}{7}})} & \ctrl{-1} & \qw & \gate{\mathrm{R_Y}\,(\mathrm{\frac{4\pi}{7}})} & \qw & \gate{\mathrm{R_Y}\,(\mathrm{\frac{2\pi}{7}})} & \gate{\mathrm{R_Y}\,(\mathrm{\frac{\pi}{7}})} & \qw & \ctrl{-1} & \gate{\mathrm{R_Z}\,(\mathrm{\theta_4})} & \qw & \gate{\mathrm{R_Y}\,(\mathrm{\theta_5})} & \qw & \qw & \qw & \qw & \qw & \qw & \qw & \qw & \qw & \qw & \qw & \targ & \ctrl{-1} &  \qw & \qw & \qw & \qw\\
        \nghost{\ket{{0}} } & \lstick{\ket{{0}} } & \qw & \qw & \qw & \qw & \targ & \qw & \qw & \qw & \qw & \qw & \qw & \qw & \targ & \qw & \qw & \qw & \qw & \targ & \gate{\mathrm{R_Y}\,(\mathrm{\theta_6})} & \qw & \gate{\mathrm{R_Z}\,(\mathrm{\frac{2\pi}{3}})} & \qw & \gate{\mathrm{R_Z}\,(\mathrm{\frac{\pi}{3}})} & \qw & \gate{\mathrm{R_Y}\,(\mathrm{\theta_7})} & \qw & \gate{\mathrm{R_Z}\,(\mathrm{\frac{4\pi}{7}})} & \gate{\mathrm{R_Z}\,(\mathrm{\frac{2\pi}{7}})} & \qw & \gate{\mathrm{R_Z}\,(\mathrm{\frac{\pi}{7}})} & \qw & \gate{\mathrm{R_Y}\,(\mathrm{\theta_8})} & \qw & \ctrl{-1} &  \qw & \qw & \qw & \qw & \qw\\
    \\ }}
    \caption{The CPQC constructed from the trained PQC \Cref{fig:european_basket_call_circuit} combined with distribution loading operators $P_1, P_2$ for the European basket call with a fixed weight.}
    \label{fig:european_basket_call_circuit_controlled}

    \resizebox{\linewidth}{!}{
        \Qcircuit @C=0.2em @R=0.3em @!R { \\
        \nghost{\ket{{0}} } & \lstick{\ket{{0}} } & \gate{\mathrm{R_X}\,(\mathrm{w_1})} & \ctrl{1} & \qw & \qw & \qw & \qw & \qw & \ctrl{1} & \gate{\mathrm{R_Y}\,(\mathrm{\theta_0})} & \gate{\mathrm{R_Z}\,(\mathrm{x_1})} & \gate{\mathrm{R_Y}\,(\mathrm{\theta_1})} & \ctrl{1} & \gate{\mathrm{R_X}\,(\mathrm{\theta_2})} & \ctrl{1} & \gate{\mathrm{R_X}\,(\mathrm{\theta_3})} & \qw & \qw & \targ & \gate{\mathrm{R_Y}\,(\mathrm{\theta_4})} & \targ & \qw & \qw & \qw & \ctrl{1} & \qw & \ctrl{1} & \gate{\mathrm{R_X}\,(\mathrm{\theta_5})} & \ctrl{1} & \qw & \ctrl{1} & \targ & \meter \\
        \nghost{\ket{{0}} } & \lstick{\ket{{0}} } & \gate{\mathrm{R_Y}\,(\mathrm{\theta_6})} & \targ & \gate{\mathrm{R_Z}\,(\mathrm{x_0})} & \gate{\mathrm{\sqrt{X}}} & \gate{\mathrm{R_Z}\,(\mathrm{\theta_7})} & \gate{\mathrm{\sqrt{X}}} & \gate{\mathrm{X}} & \targ & \gate{\mathrm{R_Y}\,(\mathrm{\theta_8})} & \qw & \qw & \targ & \gate{\mathrm{R_Z}\,(\mathrm{\theta_9})} & \targ & \gate{\mathrm{R_Z}\,(\mathrm{\theta_{10}})} & \gate{\mathrm{R_Z}\,(\mathrm{x_0})} & \gate{\mathrm{R_Y}\,(\mathrm{\theta_{11}})} & \ctrl{-1} & \qw & \ctrl{-1} & \gate{\mathrm{R_Z}\,(\mathrm{x_1})} & \gate{\mathrm{R_Y}\,(\mathrm{\theta_{12}})} & \gate{\mathrm{R_Z}\,(\mathrm{x_1})} & \targ & \gate{\mathrm{R_Z}\,(\mathrm{\theta_{13}})} & \targ & \gate{\mathrm{R_Z}\,(\mathrm{\theta_{14}})} & \targ & \gate{\mathrm{R_X}\,(\mathrm{\theta_{15}})} & \targ & \ctrl{-1} & \qw\\
    \\ }}
    \caption{Genetically trained PQC for the payoff of a European basket call with variable weight generated by \Cref{alg:structure_learning}. \(\bm{\theta}\) = [1.34793, 1.39065, 0.56594, 0.40454, -2.45389, 1.54508, 2.17214, -0.50158, -0.63069, -1.14638, -0.44283, 0.44037, -0.42722, -0.20744, -0.23896, 0.10707]}
    \label{fig:european_basket_call_variable_weight_circuit}
\end{figure}
\end{landscape}

\end{document}